\documentclass[aps,reprint,twocolumn,superscriptaddress]{revtex4-2}

\usepackage{hyperref}
\usepackage{amsmath}
\usepackage{amsfonts}
\usepackage{amssymb}
\usepackage{graphicx}
\usepackage{empheq}
\usepackage{diagbox}
\usepackage{tikz,xcolor,hyperref}

\definecolor{linkcolour}{HTML}{000066}	%light purple link for the email
\hypersetup{colorlinks,breaklinks,
	urlcolor=linkcolour, 
	linkcolor=linkcolour,
	citecolor=linkcolour}

\definecolor{lime}{HTML}{A6CE39}
\DeclareRobustCommand{\orcidicon}{
	\begin{tikzpicture}
		\draw[lime, fill=lime] (0,0) 
		circle [radius=0.16] 
		node[white] {{\fontfamily{qag}\selectfont \tiny ID}};
		\draw[white, fill=white] (-0.0625,0.095) 
		circle [radius=0.007];
	\end{tikzpicture}
	\hspace{-2mm}
}

\newcommand{\orcidauthorNB}{\href{https://orcid.org/0000-0002-1554-3820}{\orcidicon}} % Niall Byrnes
\newcommand{\orcidauthorMRF}{\href{https://orcid.org/0000-0001-5864-9636}{\orcidicon}} % Matthew Foreman

\hypersetup{colorlinks=true, urlcolor=blue, linkcolor=blue, citecolor=blue,plainpages=false} 

\newcommand{\tr}{\mbox{tr}}

\newcommand{\partialdiff}[2]{\partial_{#2}#1}

\newcommand{\res}[2]{\mathop{\mathrm{Res}}_{\omega = #1}#2}
\newcommand{\resk}[2]{\mathop{\mathrm{Res}}_{k = #1}#2}

\newcommand{\real}[1]{\mathrm{Re}#1}
\newcommand{\imag}[1]{\mathrm{Im}#1}
\newcommand{\dee}{\mathrm{d}}
\usepackage{xcolor}

\begin{document}

\preprint{APS/123-QED}

\title{Generalized Wigner-Smith analysis of resonance perturbations in {arbitrary} $Q$ non-Hermitian systems} 

\author{Niall Byrnes\orcidauthorNB}
\email[]{niallfrancis.byrnes@ntu.edu.sg}
\affiliation{School of Electrical and Electronic Engineering, Nanyang Technological University, 50 Nanyang Avenue, Singapore 639798}

\author{Matthew R. Foreman\orcidauthorMRF}
\email[]{matthew.foreman@ntu.edu.sg}
\affiliation{School of Electrical and Electronic Engineering, Nanyang Technological University, 50 Nanyang Avenue, Singapore 639798}
\affiliation{Institute for Digital Molecular Analytics and Science, 59 Nanyang Drive, Singapore 636921}

\date{\today}

\begin{abstract}
Perturbing resonant systems causes shifts in their associated scattering poles in the complex plane. In a previous study \cite{Byrnes2024b}, we demonstrated that these shifts can be calculated numerically by analyzing the residue of a generalized Wigner-Smith operator associated with the perturbation parameter. In this work, we extend this approach by connecting the Wigner-Smith formalism with results from standard electromagnetic perturbation theory applicable to open systems with resonances of arbitrary quality factors. We further demonstrate the utility of the method through several numerical examples, including the inverse design of a multi-layered nanoresonator sensor and an analysis of the enhanced sensitivity of scattering zeros to perturbations.

%\begin{description}
%Secondary publications and %information retrieval purposes.
%\end{description}
\end{abstract}

\maketitle

\section{Introduction}\label{sec:introduction} 

Optical resonances underpin a broad array of technologies and devices, including optical sensors \cite{santos2014handbook}, filters \cite{filtersbook}, tweezers \cite{Juan2011},  antennas \cite{Muhlschlegel2005}, plasmonic solar cells \cite{Catchpole2008}, metasurface holograms \cite{Huang2018}, and super-resolution imaging \cite{Willets2017}. In the design and application of such systems, it is often critical to understand how resonances are influenced by perturbations from the environment, such as from mechanical vibrations, fluctuations in humidity and temperature, material defects, and chemical contamination. For open systems, resonances manifest mathematically as complex frequency poles of quantities that characterize absorption and scattering, such as reflection coefficients and extinction cross sections \cite{Krasnok2019}. When a system is perturbed, a pole $\omega_p$ will undergo a shift $\Delta\omega_p$, altering both the line width and central frequency of the resonance. A classic first order result for the pole shift in a non-magnetic resonant cavity gives \cite{Waldron1960}
\begin{align}\label{eq:traditional}
    \frac{\Delta\omega_p}{\omega_p} = -\int_\Omega \Delta \epsilon(\mathbf{r},\omega_p)\mathbf{E}(\mathbf{r})\cdot\mathbf{E}^*(\mathbf{r})\,\dee V,
\end{align} 
where $\Delta \epsilon$ is the pointwise change in the system permittivity induced by the perturbation and the electric field $\mathbf{E}$ is that existing throughout the volume $\Omega$ of the \emph{unperturbed} system at the resonant frequency $\omega_p$. Importantly, it is implicitly assumed in Eq.~(\ref{eq:traditional}) that the fields are normalized by the stored electromagnetic energy, viz., 
\begin{align}\label{eq:energy-normalization}
\int[\epsilon(\mathbf{r},\omega_p)\mathbf{E}(\mathbf{r})\cdot\mathbf{E}^*(\mathbf{r}) + \mu_0\mathbf{H}(\mathbf{r})\cdot\mathbf{H}^*(\mathbf{r})]\,\dee V = 1,
\end{align}
where $\epsilon$ is the system permittivity, $\mu_0$ is the permeability of free space, and $\mathbf{H}$ is the magnetic field. Eq.~(\ref{eq:traditional}) therefore states that the fractional shift in the resonant frequency is determined by the ratio of the work done in polarizing the excess permittivity to the total mode energy.

Though physically appealing, Eqs.~(\ref{eq:traditional}) and (\ref{eq:energy-normalization}) suffer from numerous issues. First, Eq.~(\ref{eq:energy-normalization}) is mathematically problematic as the fields associated with leaky resonances, termed quasi-normal modes, diverge at infinity \cite{https://doi.org/10.1002/lpor.201700113}. Care is thus required in consistently normalizing such modes across different system geometries in a way that is amenable to efficient numerical computation. More pressing, however, is the fact that Eq.~(\ref{eq:traditional}) is only strictly valid for a resonance whose associated quality factor\ $Q = \real(\omega_p)/2|\imag(\omega_p)|$ satisfies $Q \gg 1$, and therefore does not apply in general \cite{https://doi.org/10.1002/lpor.201700113}. {Plasmonic nanoparticle resonators, for example, typically exhibit low $Q$ resonances for which Eq.~(\ref{eq:traditional}) is inaccurate.} Fortunately, it can be shown that Eq.~(\ref{eq:traditional}) can be modified to accurately describe resonances of arbitrary quality factors by making the modest change $\mathbf{E}\cdot\mathbf{E}^* \to \mathbf{E}\cdot\mathbf{E}$ \cite{Yang2015}. Alongside this transformation, Eq.~(\ref{eq:energy-normalization}) must also be altered so that the associated quasi-normal mode is properly normalized throughout space. Various approaches for achieving this have been proposed in the literature \cite{PhysRevA.92.053810}, each of which employ different strategies for handling the diverging fields far away from the system, such as through truncation of the integration domain \cite{Foreman2014} or the use of perfectly matched layers \cite{PhysRevLett.110.237401}. 

{A common approach to the study of non-Hermitian systems more generally is the effective Hamiltonian, where loss and coupling to the environment are treated as perturbations to an underlying Hermitian Hamiltonian \cite{Ashida02072020}. Alternatively, the frequency domain scattering matrix $\mathbf{S}(\omega)$ allows for the description of open systems through consideration of fields that scatter away from it} \cite{RevModPhys.89.015005}. As with other scattering quantities, the scattering matrix can be analytically continued to a meromorphic function whose poles correspond to scattering resonances. We recently proved that a perturbation-induced complex resonance shift can be expressed in terms of the residue of an associated generalized Wigner-Smith operator $\mathbf{Q}_\xi = -i\mathbf{S}^{-1}\partialdiff{\mathbf{S}}{\xi}$ (not to be confused with $Q$), viz.,
\begin{align}\label{eq:ws-formula}
    \Delta\omega_p =  -\Delta\alpha \frac{\res{\omega_p}{\tr(\mathbf{Q}_\alpha)}}{\res{\omega_p}{\tr(\mathbf{Q}_\omega)}} = i\Delta\alpha \res{\omega_p}{\tr(\mathbf{Q}_\alpha)},
\end{align}
where $\alpha$ is a generic system parameter that is shifted by $\Delta\alpha$ in response to the perturbation, $\mathrm{tr}$ denotes the trace operator, and $\mathrm{Res}$ denotes the complex residue \cite{Byrnes2024b}. {Unlike Eq.~(\ref{eq:traditional}), Eq.~(\ref{eq:ws-formula}) is valid for arbitrary values of $Q$. In addition, since the Wigner-Smith operator is based on the scattering matrix, resonance shifts can be calculated without consideration of the system's internal fields. We also note that} although the current work focuses on optical and plasmonic resonances, Eq.~(\ref{eq:ws-formula}) was derived from complex analytic arguments and is thus applicable more generally to other domains of resonant physics, including for example, particle physics \cite{creque2021resonant}, acoustic engineering \cite{CHOI2022104907} and cosmology \cite{PhysRevLett.100.141301}.

In Ref.~\cite{Byrnes2024b}, we demonstrated that Eq.~(\ref{eq:traditional}) can be derived from Eq.~(\ref{eq:ws-formula}) {as a special case} in the limit $Q \gg 1$. {In this work, we extend our theory to systems with arbitrary $Q$ and show that the discussed modifications to Eqs.~(\ref{eq:traditional}) and (\ref{eq:energy-normalization}) can also be derived from Eq.~(\ref{eq:ws-formula}), further highlighting the flexibility of our approach. Furthermore,} we present additional generalizations, examples and applications of Eq.~(\ref{eq:ws-formula}), extending the results of Ref.~\cite{Byrnes2024b}. In section \ref{sec:theory}, we review the key results of Ref. \cite{Byrnes2024b} and show how the theory presented within can be generalized to arbitrary $Q$ factors. We also discuss how the theory can be applied to open scattering geometries, more general functions associated with absorption and scattering, and shifts in scattering zeros, which can be assoicated with, for example, coherent perfect absorption modes \cite{cpa}. In section \ref{sec:examples}, we apply our theory to several practical examples, including a homogeneous sphere and a multi-layered spherical nano-resonator. In the latter case we consider the inverse design problem of optimizing the resonator's sensitivity to perturbations in the refractive index of the surrounding medium. Finally, we briefly compare the sensitivity of shifts in scattering poles and zeros before summarizing our work in section \ref{sec:conclusion}.

\section{Theory}\label{sec:theory}

\subsection{Perturbation theory}\label{sec:perturbation}
The derivation of Eq.~(\ref{eq:traditional}) from Eq.~(\ref{eq:ws-formula}) was facilitated by the fact that the generalized Wigner-Smith operator admits the factorization $\mathbf{Q}_{\xi} = \mathbf{A}^{-1}\mathbf{B}_\xi$, where $\mathbf{A} = \mathbf{S}^\dagger \mathbf{S}$ and $\mathbf{B}_\xi = -i\mathbf{S}^{\dagger}\partialdiff{\mathbf{S}}{\xi}$. Note that $\mathbf{B}_\xi$ coincides with the usual definition of the Wigner-Smith matrix at real frequencies, since $\mathbf{A}$ is equal to the identity matrix for lossless systems \cite{Byrnes2021a}. For a system coupled to the environment by a collection of waveguides, energy balance considerations lead to
\begin{align}
    \mathbf{A} &= \mathbf{I} + 2\imag(\omega)\int_\Omega(\epsilon\mathbf{U}^{e} + \mu_0\mathbf{U}^{m})\,\mathrm{d}V,\label{eq:A} \\
    \mathbf{B}_\xi &= -\int_\Omega[\partialdiff{(\omega\epsilon)}{\xi}\mathbf{U}^{e} +\partialdiff{\omega}{\xi}\mu_0\mathbf{U}^{m} \nonumber \\
    &\quad\quad\quad\quad+ 2i\imag(\omega)(\epsilon\mathbf{V}_\xi^{e} +\mu_0\mathbf{V}_\xi^{m})]\,\dee V,\label{eq:B}
\end{align}
where $\mathbf{I}$ is the identity matrix and $\mathbf{U}^e, \mathbf{U}^m, \mathbf{V}^e$ and $\mathbf{V}^m$ are matrices whose elements involve different quadratic forms of the electric and magnetic fields, as well as their derivatives, distributed throughout the system when it is excited by different incident modes \cite{Byrnes2024b}. Implicit in Eqs.~(\ref{eq:A}) and (\ref{eq:B}) is the fact that the modes in the connecting waveguides satisfy the orthonormality relation
\begin{align}\label{eq:norm-conj}
\frac{1}{2}\int_{\partial\Omega_m}\mathbf{e}_{p,t} \times \mathbf{h}^*_{q,t} \cdot \hat{\mathbf{n}}\,\dee A = \delta_{pq},
\end{align}
where $\delta_{pq}$ is the Kronecker delta, and $\mathbf{e}_{p,t}$ and $\mathbf{h}_{q,t}$ are the transverse parts of the electric and magnetic fields of the $p$'th and $q$'th waveguide mode profiles respectively. We assume that these modes exist within the $m$`th waveguide with cross section $\partial\Omega_m$ and unit normal vector $\hat{\mathbf{n}}$. Eq.~(\ref{eq:norm-conj}) follows from the reciprocity theorem and is valid for \emph{lossless} waveguides \cite{snyder2012optical}. Since we are concerned with resonances, however, it is necessary to consider the case where the system contains fields that oscillate at \emph{complex} frequencies, whereby, even if the system is lossless in the sense that $\imag{(\epsilon)} = 0$, the non-zero value of $\imag{(\omega)}$ will introduce effective loss or gain to the system. This problem can be neglected in the case $Q \gg 1$, whereby the resonant mode only couples weakly to the environment and Eq.~(\ref{eq:norm-conj}) thus holds approximately. A more satisfying resolution, however, is to note that the waveguide modes always satisfy the more general reciprocity relation
\begin{align}\label{eq:unconj}
\frac{1}{2}\int_{\partial\Omega_m}\mathbf{e}_{p,t} \times \mathbf{h}_{q,t} \cdot \hat{\mathbf{n}}\,\dee A = \delta_{pq},
\end{align}
irrespective of loss or gain present in the system \cite{snyder2012optical}. With this in mind, we need not make assumptions on the value of $Q$ if we instead express $\mathbf{Q}_\xi$ in terms of \emph{unconjugated} field products. This motivates the alternate factorization $\mathbf{Q}_{\xi} = \mathbf{C}^{-1}\mathbf{D}_\xi$, where $\mathbf{C} = \mathbf{S}^\mathrm{T} \mathbf{S}$ and $\mathbf{D}_\xi = -i\mathbf{S}^{\mathrm{T}}\partialdiff{\mathbf{S}}{\xi}$. Note that $\mathbf{C}$ and $\mathbf{D}_\xi$ are formally the same as $\mathbf{A}$ and $\mathbf{B}_\xi$, except the adjoint operators ($\dagger$) are replaced with transpose operators ($\mathrm{T}$). By deriving equations for $\mathbf{C}$ and $\mathbf{D}_\xi$ analogous to Eqs. (\ref{eq:A}) and (\ref{eq:B}), we are able to express $\mathbf{Q}_\xi$ in terms of unconjugated field products, which finally leads to the generalization of Eq.~(\ref{eq:traditional}) discussed in section \ref{sec:introduction}. We now outline the details of this calculation. 

To start, we assume that the fields are time harmonic and satisfy Maxwell's equations
\begin{align}
    \nabla \times \mathbf{E} &= i\omega\mu_0 \mathbf{H},\\
    \nabla \times \mathbf{H} &= -i\omega\epsilon \mathbf{E},
\end{align}
where both $\epsilon$ and $\omega$ may be complex. Following the usual derivation of the Poynting theorem \cite{Geyi:19}, it is straightforward to show that
\begin{align}\label{eq:poynt}
\begin{split}
    -\frac{i}{4}\int_{\partial\Omega}&\mathbf{E}_p\times\mathbf{H}_q \cdot \hat{\mathbf{n}}\,\dee A\\
    &= \frac{\omega}{4}\int_\Omega(\epsilon\mathbf{E}_p\cdot\mathbf{E}_q + \mu_0\mathbf{H}_p\cdot\mathbf{H}_q)\,\dee V,
\end{split}
\end{align}
where $\Omega$ is an arbitrary volume with boundary $\partial\Omega$ whose local unit normal vector is $\hat{\mathbf{n}}$. The subscripts $p$ and $q$ in Eq.~(\ref{eq:poynt}) enumerate all waveguide modes and indicate that the fields are those that exist throughout the system when it is illuminated by the corresponding mode. We assume that the surface $\partial\Omega$ fully encapsulates the system and is perforated only by the waveguides, which, for the sake of mathematical simplicity, we take to each be single mode. The transverse parts of the fields within the $m$'th waveguide that arise due to illuminating the $p$'th waveguide are given by
\begin{align}
	\mathbf{E}_{mp,t} &= (\delta_{mp}e^{-i\beta_m z} + S_{mp}e^{i\beta_m z})\mathbf{e}_{m,t},\label{eq:E-guide}\\
	\mathbf{H}_{mp,t} &= (\delta_{mp}e^{-i\beta_m z} - S_{mp}e^{i\beta_m z})\mathbf{h}_{m,t},\label{eq:H-guide}
\end{align}
where $z$ is a local coordinate along the waveguide axis such that $z=0$ on $\partial\Omega$, $S_{mp}$ is an element of the system scattering matrix, and $\beta_m$ is the mode propagation constant. Inserting Eqs.~(\ref{eq:E-guide}) and (\ref{eq:H-guide}) into the left hand side of Eq.~(\ref{eq:poynt}) and using Eq.~(\ref{eq:unconj}) leads to the result
\begin{align}\label{eq:C}
    \mathbf{C} = \mathbf{I} - 2i\omega\int_\Omega(\epsilon \widetilde{\mathbf{U}}^{e} + \mu_0\widetilde{\mathbf{U}}^{m})\,\dee V,
\end{align}
where $\widetilde{\mathbf{U}}^e$ and $\widetilde{\mathbf{U}}^m$ are matrices with elements $\widetilde{U}^e_{qp} = \mathbf{E}_p\cdot\mathbf{E}_q/4$ and $\widetilde{U}^m_{qp} = \mathbf{H}_p\cdot\mathbf{H}_q/4$. Note that Eq.~(\ref{eq:C}) does not yield an analogue to the usual unitarity relation $\mathbf{A} = \mathbf{S}^\dagger\mathbf{S} = \mathbf{I}$ at real frequencies, since the second term on the right hand side of Eq.~(\ref{eq:C}) is generally non-zero for all values of $\omega$. 

To obtain an integral expression for $\mathbf{D}_\xi$, we begin with the equation
\begin{align}\label{eq:energy2}
\begin{split}
    -\frac{i}{4}&\int_{\partial\Omega}(\mathbf{E}_p\times\partialdiff{\mathbf{H}_q}{\xi} \pm \partialdiff{\mathbf{E}_p}{\xi}\times\mathbf{H}_q) \cdot \hat{\mathbf{n}}\,\dee A\\
    = &\frac{1}{4}\int_\Omega[\partialdiff{(\omega\epsilon)}{\xi}\mathbf{E}_p\cdot \mathbf{E}_q \pm \partialdiff{(\omega\mu_0)}{\xi}\mathbf{H}_p\cdot\mathbf{H}_q]\,\dee V\\
    &+ \frac{\omega}{4}\int_\Omega\epsilon(\mathbf{E}_p\cdot \partialdiff{\mathbf{E}_q}{\xi} \pm \partialdiff{\mathbf{E}_p}{\xi}\cdot\mathbf{E}_q)\,\dee V\\
    &+ \frac{\omega}{4}\int_\Omega\mu_0(\mathbf{H}_p\cdot \partialdiff{\mathbf{H}_q}{\xi} \pm \partialdiff{\mathbf{H}_p}{\xi}\cdot\mathbf{H}_q)\,\dee V,
    \end{split}
\end{align}
which follows from Maxwell's equations and applying the divergence theorem to the vector identity
\begin{align}
\nabla\cdot(\mathbf{E}\times\partialdiff{\mathbf{H}}{\xi}) = \partialdiff{\mathbf{H}}{\xi}\cdot\nabla\times\mathbf{E}  - \partialdiff{\mathbf{E}}{\xi}\cdot\nabla\times\mathbf{H}.  
\end{align}
Eq.~(\ref{eq:energy2}) is in fact a pair of vector equations, since one may choose either the upper or lower set of plus and minus signs.
Using the field expressions in Eqs.~(\ref{eq:E-guide}) and (\ref{eq:H-guide}) and their derivatives, the upper and lower set of plus and minus signs of Eq.~(\ref{eq:energy2}) give, respectively,
\begin{align}\label{eq:upper}
\begin{split}
    \mathrm{Sym}&(\mathbf{D}_\xi)\\
    &= \int_\Omega [\partialdiff{(\omega\epsilon)}{\xi}\widetilde{\mathbf{U}}^e + \partialdiff{(\omega\mu_0)}{\xi}\widetilde{\mathbf{U}}^m\\
    &\quad\quad+ \omega(\epsilon\mathrm{Sym}(\widetilde{\mathbf{V}}^e_\xi) + \mu_0\mathrm{Sym}(\widetilde{\mathbf{V}}^m_\xi))]\,\dee V,
    \end{split}
\end{align}
\begin{align}
\begin{split}
    \mathrm{ASym}&(\mathbf{D}_\xi) -i\mathrm{Sym}(\partialdiff{\mathbf{S}}{\xi})\\
    &= \int_\Omega [\partialdiff{(\omega\epsilon)}{\xi}\widetilde{\mathbf{U}}^e - \partialdiff{(\omega\mu_0)}{\xi}\widetilde{\mathbf{U}}^m\\&\quad\quad+\omega(\epsilon\mathrm{ASym}(\widetilde{\mathbf{V}}^e_\xi) + \mu_0\mathrm{ASym}(\widetilde{\mathbf{V}}^m_\xi))]\,\dee V,\label{eq:lower}
\end{split}
\end{align}
where $\widetilde{\mathbf{V}}^e$ and $\widetilde{\mathbf{V}}^m$ are matrices with elements $\widetilde{V}^e_{qp} = \partialdiff{\mathbf{E}_p}{\xi}\cdot\mathbf{E}_q/4$ and $\widetilde{V}^m_{qp} = \partialdiff{\mathbf{H}_p}{\xi}\cdot\mathbf{H}_q/4$, and $\mathrm{Sym}$ and $\mathrm{ASym}$ denote the symmetric and anti-symmetric parts of a matrix, defined by $\mathrm{Sym}(\mathbf{M}) = (\mathbf{M} + \mathbf{M}^\mathrm{T})/2$ and $\mathrm{ASym}(\mathbf{M}) = (\mathbf{M} - \mathbf{M}^\mathrm{T})/2$ respectively. Since $\mathbf{D}_\xi = \mathrm{Sym}(\mathbf{D}_\xi) + \mathrm{ASym}(\mathbf{D}_\xi)$, we can add Eq.~(\ref{eq:upper}) to the anti-symmetric part of both sides of Eq.~(\ref{eq:lower}) to obtain
\begin{align}\label{eq:Dunconj}
    \mathbf{D}_\xi = \int_\Omega [\partialdiff{(\omega\epsilon)}{\xi}\widetilde{\mathbf{U}}^e + \partialdiff{\omega}{\xi}\mu_0\widetilde{\mathbf{U}}^m + \omega(\epsilon\widetilde{\mathbf{V}}^e_\xi + \mu_0\widetilde{\mathbf{V}}^m_\xi)]\,\dee V.
\end{align}
Finally, since $\mathbf{Q}_\xi = \mathbf{C}^{-1}\mathbf{D}_\xi$, combining Eqs.~(\ref{eq:C}) and (\ref{eq:Dunconj}) gives us an expression for $\mathbf{Q}_\xi$ in terms of the system's internal fields. We emphasize that, in contrast to Eqs.~(\ref{eq:A}) and (\ref{eq:B}) for $\mathbf{A}$ and $\mathbf{B}_\xi$, Eqs.~(\ref{eq:C}) and (\ref{eq:Dunconj}) cannot be interpreted as energy balance relations. Note, for example, that the diagonal elements of $\widetilde{\mathbf{U}}^e$ and $\widetilde{\mathbf{U}}^m$ are generally complex valued and thus cannot correspond to traditional energy densities. Nevertheless, Eqs.~(\ref{eq:C}) and (\ref{eq:Dunconj}) are more general than Eqs.~(\ref{eq:A}) and (\ref{eq:B}) and hold for arbitrary values of $Q$.

Armed with expressions for $\mathbf{Q}_\xi$, we now return to interpreting Eq.~(\ref{eq:ws-formula}). If, for simplicity, we assume that the system is coupled to the environment by a single waveguide mode, then $\mathbf{C}$ and $\mathbf{D}_\xi$ reduce to scalar quantities $C$ and $D_\xi$. Using the limit definition of residues and the first equality of Eq.~(\ref{eq:ws-formula}), it can be seen that a resonant frequency $\omega_p$ will shift in response to a perturbation $\Delta \alpha$ according to
\begin{align}\label{eq:res-lim}
    \Delta\omega_p \approx -\Delta\alpha \frac{C^{-1}D_\alpha}{C^{-1}D_\omega} = -\Delta\alpha \frac{D_\alpha}{D_\omega},
\end{align}
where the right hand side of Eq.~(\ref{eq:res-lim}) should be evaluated at a frequency close to $\omega_p$. We assume without loss of generality that the fields are normalized such that $D_\omega = 1$. In addition, for a lowest order perturbation theory, we may take $\mathbf{V}^e_\alpha = \mathbf{V}^m_\alpha \approx 0$ in Eq.~(\ref{eq:Dunconj}), implying that the internal fields only vary weakly with the perturbation parameter.
Under this assumption, the perturbation only serves to modify $\epsilon$ within the system, and noting $\Delta \alpha \partialdiff{\epsilon}{\alpha} \approx \Delta\epsilon$, Eq.~(\ref{eq:res-lim}) finally leads to
\begin{align}\label{eq:unconjresult}
    \frac{\Delta\omega_p}{\omega_p} = -\int_{\Omega}\Delta\epsilon(\mathbf{r},\omega_p)\mathbf{E}(\mathbf{r})\cdot\mathbf{E}(\mathbf{r})\,\dee V,
\end{align}
which is precisely the modified form of Eq.~(\ref{eq:traditional}) originally sought. We note that our asserting $D_\omega = 1$ is equivalent to normalizing the resonance's quasi-normal mode. The right hand side of Eq.~(\ref{eq:ws-formula}), however, does not involve $\mathbf{Q}_\omega$, since it can be shown that $\res{\omega_p}{\tr(\mathbf{Q}_\omega)} = i$ \cite{Byrnes2024b}. This suggests that the problematic infinities associated with quasi-normal modes are dealt with by the residues in Eq.~(\ref{eq:ws-formula}). Eq.~(\ref{eq:ws-formula}) can therefore be used without particular consideration of mode normalization, provided that $\mathbf{S}$ (and therefore $\mathbf{Q}_\xi$) is known.

We have shown that, starting from Eq.~(\ref{eq:ws-formula}), it is possible to derive both Eqs.~(\ref{eq:traditional}) and (\ref{eq:unconjresult}) using a suitable factorization of $\mathbf{Q}_\xi$. For the remainder of section \ref{sec:theory}, we shall discuss other generalizations and useful modifications of Eq.~(\ref{eq:ws-formula}).

\subsection{Wavenumber perturbations}
The resonance shift in Eq.~(\ref{eq:ws-formula}) is expressed as a change in frequency. In some cases, however, it may be more convenient to express the resonance shift as a change in wavenumber. Although wavenumber generally varies throughout a system as a result of spatial heterogeneity, in many practical cases the medium external to system will be air, vacuum, or some other non-dispersive medium. In such cases, the wavenumber in the background medium is given by $k = \omega / c$, where $c$ is the speed of light in the background medium. A scattering matrix pole $\omega_p$ therefore corresponds to a background complex wavenumber $k_p = \omega_p/c$. It is simple to show that Eq.~(\ref{eq:ws-formula}) can also be used to determine the shift in wavenumber $\Delta k_p = \Delta \omega_p /c$ by replacing all instances of $\omega$ in Eq.~(\ref{eq:ws-formula}) with $k$. Naturally, the residues should then be taken in the complex $k$ plane. To avoid confusion, for the remainder of this work it shall be assumed that $k$ refers to the wavenumber \emph{in vacuum}. Refractive indices, where necessary, shall always be expressed explicitly.

\subsection{Open geometries}
In section \ref{sec:perturbation}, it was assumed that the system was coupled to the environment by a collection of waveguides. This influenced the expressions for the fields on $\partial\Omega$, as expressed in Eqs.~(\ref{eq:E-guide}) and (\ref{eq:H-guide}), as well as the mode normalization given by Eq.~(\ref{eq:unconj}). Although clearly not applicable to all systems, these assumptions are not as restrictive as they may first appear. Many scattering problems involve open geometries, where scattered fields directly radiate away from the system in free space. In such cases, it is more appropriate to describe the fields using free space modes. We could, for example, write the fields far away from the system as \cite{mishchenko2002scattering}
\begin{align}
	\mathbf{E}_{p,t} &= \sum_q\bigg[\delta_{qp}\frac{e^{-ik(r-R)}}{kr} + S_{qp}\frac{e^{ik(r-R)}}{kr}\bigg]\mathbf{e}_{q,t},\label{eq:E-open}\\
	\mathbf{H}_{p,t} &= \sum_q\bigg[\delta_{qp}\frac{e^{-ik(r-R)}}{kr} - S_{qp}\frac{e^{ik(r-R)}}{kr}\bigg]\mathbf{h}_{q,t},\label{eq:H-open}
\end{align}
where $r$ is the radial coordinate of a spherical polar coordinate system centred on the system and $R$ is the radius of a large reference sphere. In Eqs.~(\ref{eq:E-open}) and (\ref{eq:H-open}), $\mathbf{e}_{q,t}$ and $\mathbf{h}_{q,t}$ describe the transverse parts of suitable vector spherical harmonics, which are functions of the polar and azimuthal coordinates, and, similarly to before, we have assumed that the system is illuminated by the $p$'th incident mode. The index $q$ in Eqs.~(\ref{eq:E-open}) and (\ref{eq:H-open}) spans different multipole orders and polarizations and, unlike the waveguide geometry, each mode exists throughout the entire background medium, rather than being spatially confined to a particular waveguide.

Using Eqs.~(\ref{eq:E-open}) and (\ref{eq:H-open}), the perturbation theory in section \ref{sec:perturbation} can be repeated in largely the same manner as before for the open geometry. We now assume, however, that $\mathbf{e}_{p,t}$ and $\mathbf{h}_{q,t}$ satisfy an equation analogous to Eq.~(\ref{eq:unconj}), but with the integration domain taken to be the surface of the reference sphere. One notable difference between the waveguide and open geometries is that, while for the waveguide geometry we have, on differentiating Eq.~(\ref{eq:E-guide}), $\partialdiff{\exp(\pm i\beta z)}{k}\,|_{z=0} = 0$, for the open geometry we instead find on differentiating Eq.~(\ref{eq:E-open}) that $\partialdiff{\exp[\pm ik(r-R)]/kr}{k}\,|_{r=R} = -1/k^2R$, which is non-zero. This means that $\mathbf{D}_k$ (or $\mathbf{D}_\omega$) will contain an additional term that is not present for the waveguide geometry. This additional term, however, does not affect $\mathbf{D}_\alpha$ and only influences the normalization of $\mathbf{D}_k$ (or $\mathbf{D}_\omega$) \cite{9142355}. 

\subsection{Alternate expressions and scattering zeros}\label{sec:alternate}
The derivation of Eq.~(\ref{eq:ws-formula}) does not make use of any particular physical properties of the scattering matrix other than the fact that it has a pole at the resonant frequency. Therefore, although the scattering matrix was useful in the perturbation theory presented in section \ref{sec:perturbation}, it is not strictly necessary to predict resonance shifts. In fact, we can replace $\mathbf{S}$ with \emph{any} other matrix function $\mathbf{M}$ that possesses the same pole landscape as $\mathbf{S}$ and use its corresponding logarithmic derivative operator $\mathbf{L}_\xi = -i\mathbf{M}^{-1}\partialdiff{\mathbf{M}}{\xi}$ in place of $\mathbf{Q}_\xi$ in Eq.~(\ref{eq:ws-formula}). This may be particularly useful in cases where the scattering matrix is not known, but one instead has some other mathematical constraint for the system's resonances, or for systems where a scattering matrix is inappropriate, e.g., closed systems, or systems with quasi-static fields. For example, a resonance condition may be of the form $\det[\boldsymbol{\Lambda}(\omega_p, \alpha)] = 0$ for some matrix $\boldsymbol{\Lambda}$. In this case, we can set $\mathbf{M} = \boldsymbol{\Lambda}^{-1}$, since $\det(\boldsymbol{\Lambda}^{-1})$ possesses at pole at $\omega_p$. By tracking shifts in the poles of $\det(\boldsymbol{\Lambda}^{-1})$, we are effectively tracking shifts in the zeros of $\det(\boldsymbol{\Lambda})$, which describe how the resonance condition is affected by the perturbation. Note that, by the exact same reasoning, in addition to tracking resonance shifts, we can also use Eq.~(\ref{eq:ws-formula}) to track shifts in a system's zeros. If, for example, the scattering matrix possesses a zero at $\omega_z$, then this will manifest as a pole of the inverse matrix $\mathbf{S}^{-1}$. Setting $\mathbf{M} = \mathbf{S}^{-1}$ and tracking the poles of $\mathbf{S}^{-1}$ is thus equivalent to tracking the zeros of $\mathbf{S}$. It should be noted that since Eq.~(\ref{eq:ws-formula}) assumes that $\mathbf{S}$ has \emph{simple} poles, it cannot be used to track a pole or zero of order or multiplicity greater than one. This case may arise, for example, in the study of exceptional points, which can correspond to higher order poles and branch points \cite{PhysRevE.67.026204}. Such considerations, however, are beyond the scope of this work.

To further illustrate the flexibility of Eq.~(\ref{eq:ws-formula}) with a specific example, consider the case of scattering by an arbitrary, isolated particle in an open, free-space geometry. As can be seen in Eqs.~(\ref{eq:E-open}) and (\ref{eq:H-open}), the scattering matrix relates the coefficients of outgoing and incoming spherical waves with different transverse wave functions. These spherical waves are the far field limiting forms of the spherical Hankel functions $h^{(1)}_\nu(kr)$ and $h^{(2)}_\nu(kr)$ respectively, where $\nu$ denotes the multipole order. It is sometimes simpler, however, to express the incident field using the spherical Bessel function $j_\nu(kr)$, which is finite at $r=0$. In this case, the relationship between the incident and scattered fields is described by the so-called $T$-matrix $\mathbf{T}$, which is related to the scattering matrix by $\mathbf{T} = (\mathbf
S - \mathbf{I})/2$ \cite{mishchenko2002scattering}. If the scatterer is a homogeneous, isotropic sphere, there is no mixing between different multipole orders, and $\mathbf{S}$ and $\mathbf{T}$ are both diagonal matrices. The elements of $\mathbf{T}$ are, up to trivial scaling, given by the usual Mie coefficients. For a sphere of radius $r_c$ and refractive index $n_c$, located in a background medium of refractive index $n_b$, the Mie coefficient for electric multipoles is given by \cite{bohren1983absorption}
\begin{align}\label{eq:mie}
    a_\nu = \frac{m\psi_\nu(mx)\psi'_\nu(x)-\psi_\nu(x)\psi_\nu'(mx)}{m\psi_\nu(mx)\xi'_\nu(x)-\xi_\nu(x)\psi_\nu'(mx)},
\end{align}
where $x=kn_br_c$ is the scatterer size parameter, $m = n_c/n_b$ is the relative refractive index and $\psi_\nu$, $\xi_\nu$ and their derivatives $\psi'_\nu$ and $\xi'_\nu$ are the Ricatti-Bessel functions. Clearly, for a scattering resonance $k_p$ occurring at a particular multipole order $\nu$, $\mathbf{S}$, $\mathbf{T}$ and $a_\nu$, all of which are linearly related, will have poles at $k_p$ and, in the context of the preceding discussion, can all be used as $\mathbf{M}$ (a scalar quantity $M$ in the case $M = a_\nu$). For more general scatterers, since there are infinite multipole orders, $\mathbf{S}$ and $\mathbf{T}$ will be of infinite size. It may be the case however, for example due to particular symmetries, that certain scattering modes do not mix, allowing the matrices to be written as a collection of decoupled blocks. One can then truncate the matrix in question to the block that contains the mode with the resonance of interest. If mixing occurs between many scattering modes, as is the case for anistropic scatterers, then in practice the matrices must be truncated by other means, such as by neglecting high order modes.  

As a final point regarding the flexibility of the choice of $\mathbf{M}$, take now $M=a_\nu$ and note that $M$ is the ratio of two functions $f$ and $g$, corresponding to the numerator and denominator of Eq.~(\ref{eq:mie}). It is possible to simplify matters further if, for example, we assume that a pole $k_p$ in $M$ arises from a zero of $g$, rather than a pole of $f$. In this case, we can also define $M' = 1/g$, which clearly also has a pole at $k_p$. Note now, however, that, by the properties of logarithmic derivatives,
\begin{align}
    M^{-1}\partialdiff{M}{\alpha} = f^{-1}\partialdiff{f}{\alpha} - g^{-1}\partialdiff{g}{\alpha}.
\end{align}
Since $f$ does not have a pole at $k_p$ (and assuming it is also non-zero at $k_p$), $f^{-1}\partialdiff{f}{\alpha}$ has no residue at $k_p$. The residue of $M^{-1}\partialdiff{M}{\alpha}$ is therefore equal to the residue of $- g^{-1}\partialdiff{g}{\alpha}$, which is also equal to the residue of $M'$. Therefore, using either $M$ or $M'$ in Eq.~(\ref{eq:ws-formula}) both yield the same pole shift. There is thus no loss in information in neglecting multiplicative factors that do not possess zeros or poles at resonant frequencies. 

\section{Examples and applications}\label{sec:examples}

In this section, we now apply Eq.~(\ref{eq:ws-formula}) to study resonance shifts in more concrete examples. Consider first a homogeneous sphere as described in section \ref{sec:alternate}, assumed to possess a pole in $a_\nu$ at wavenumber $k_p$ for a particular multipole order $\nu$. We can use Eq.~(\ref{eq:ws-formula}) to determine the resonance shift due to a small change in the radius $\Delta r_c$. A simple analytic result exists in the case that the sphere is non-dispersive, i.e. when $m$ is independent of $k$. In this case, on observing Eq.~(\ref{eq:mie}) it should be noted that $a_\nu$ depends on $k$ and $r_c$ only through the symmetric product $kr_c$. It follows that if $r_c$ changes to $r'_c$, the pole must shift to $k'_p$ so that the quantity $C = k_pr_c = k'_pr'_c$ remains constant. The shifted pole is hence given by $k'_p = k_p r_c/r'_c$. To see how this emerges from our theory, note that, by the chain rule,
\begin{align}
    &\partialdiff{a_\nu}{r_c}= \partialdiff{a_\nu}{kr_c}\partialdiff{(kr_c)}{r_c} = k\partialdiff{a_\nu}{kr_c},\label{eq:diff1}\\
    &\partialdiff{a_\nu}{k}= \partialdiff{a_\nu}{kr_c}\partialdiff{(kr_c)}{k} = r_c\partialdiff{a_\nu}{kr_c},\label{eq:diff2}
\end{align}
and thus, using Eq.~(\ref{eq:ws-formula}) and the limit definition of residues,
\begin{align}\label{eq:linear-formula}
-\frac{\Delta k_p}{\Delta r_c}= \lim_{k \to k_p}\bigg(\frac{a_\nu^{-1}\partialdiff{a_\nu}{r_c}}{a_\nu^{-1}\partialdiff{a_\nu}{k}}\bigg) = \lim_{k\to k_p}\bigg(\frac{k}{r_c}\bigg)=\frac{k_p}{r_c}.
\end{align}
Eq.~(\ref{eq:linear-formula}) shows that the sensitivity of the resonance to small radial perturbations is given by $-k_p/r_c$. In the limit of weak perturbations, Eq.~(\ref{eq:linear-formula}), can be thought of as a differential equation by making the replacement $\Delta k_p / \Delta r_c \to \mathrm{d}k_p/\mathrm{d}r_c$. This equation can be solved for $k_p$ as a function of $r_c$, which describes the trajectory of the pole in the complex plane when $r_c$ is perturbed continuously through a range of values. The solution is the rectangular hyperbola $k_pr_c = C$, where $C$ is an integration constant given by the product $k_p r_c$ before the perturbation, which fully agrees with our previous analysis.
If the sphere is dispersive so that $m$ also varies with $k$, as is often the case in reality, then poles will traverse more complicated paths in the complex plane.

\begin{figure}[t]
	\centering
\includegraphics[width=\columnwidth]{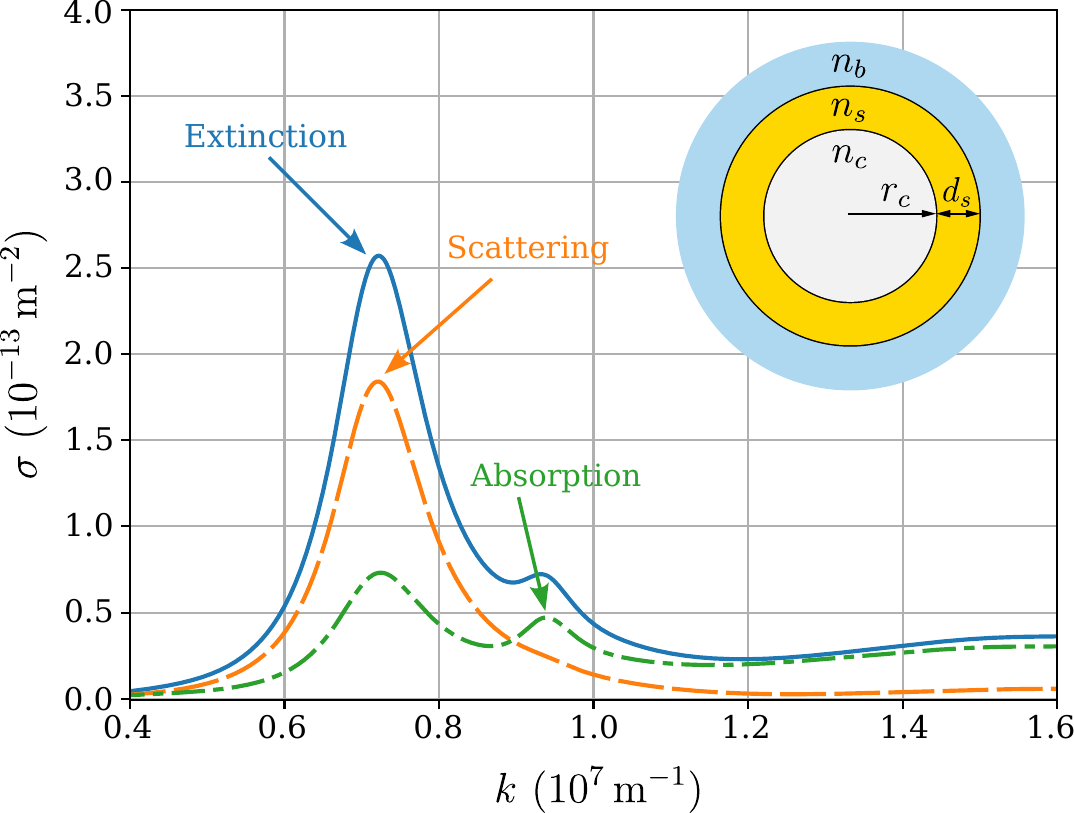}
	\caption{Extinction (blue, solid), scattering (orange, dashed) and absorption (green, dot-dashed) cross sections for a silica-gold nano-sphere in water with $r_c = 60\,\mathrm{nm}$ and $d_s = 10\,\mathrm{nm}$ as a function of $k$. A schematic diagram of the nanoparticle's multi-layer structure is also shown.}
	\label{fig:cs}
\end{figure}
In addition to tracking poles, Eq.~(\ref{eq:ws-formula}) can also be useful in inverse design problems in open, lossy systems. Consider, for example, the problem of detecting variations in the ambient refractive index of a liquid medium using a spherical, plasmonic nano-resonator \cite{Piliarik:11}. These variations might be caused by changes in the chemical composition of the surrounding environment, for example from the introduction of foreign agents. It is well known that multi-layered spheres composed of different materials can provide enhanced sensing capabilities compared to homogeneous spheres due to field enhancements concentrated near the external interface of the resonator \cite{Zhang2021}. It is therefore important to explore the parameter space of such devices to identify optimal sensing configurations. One commonly used figure of merit in assessing sensing performance is the ratio of the resonance sensitivity to the resonance width \cite{TUERSUN2016250}. Since the resonance width is proportional to $|\imag(k_p)|$, we can define the complex metric $\eta_{b}$ by 
\begin{align}\label{eq:fom}
    \eta_{b} = \frac{\Delta k_p}{|\imag(k_p)|} = \frac{i\resk{k_p}{L_{n_b}}}{|\imag{(k_p)}|},
\end{align}
where $L_{n_b} = -iM^{-1}\partialdiff{M}{n_b}$ is a logarithmic derivative operator associated with any scattering function $M$ in which the resonance manifests. The real and imaginary parts of $\eta_{b}$ yield a pair of metrics. $\real(\eta_{b})$ is equal to $\real(\Delta k_p)/|\imag(k_p)|$ and is thus associated with the shift of the central resonant frequency. This metric discredits resonances for which the frequency shift is small relative to the line width. In practice, such resonance shifts can be difficult to detect in the presence of noise \cite{Foreman2014a}. Similarly, $\imag(\eta_{b})$ is equal to $\imag(\Delta k_p)/|\imag(k_p)|$ and is thus associated with the broadening or narrowing of the resonance, which can also be used as a sensing mechanism \cite{Shao2013}.
 
\begin{figure}[t]
	\centering
\includegraphics[width=\columnwidth]{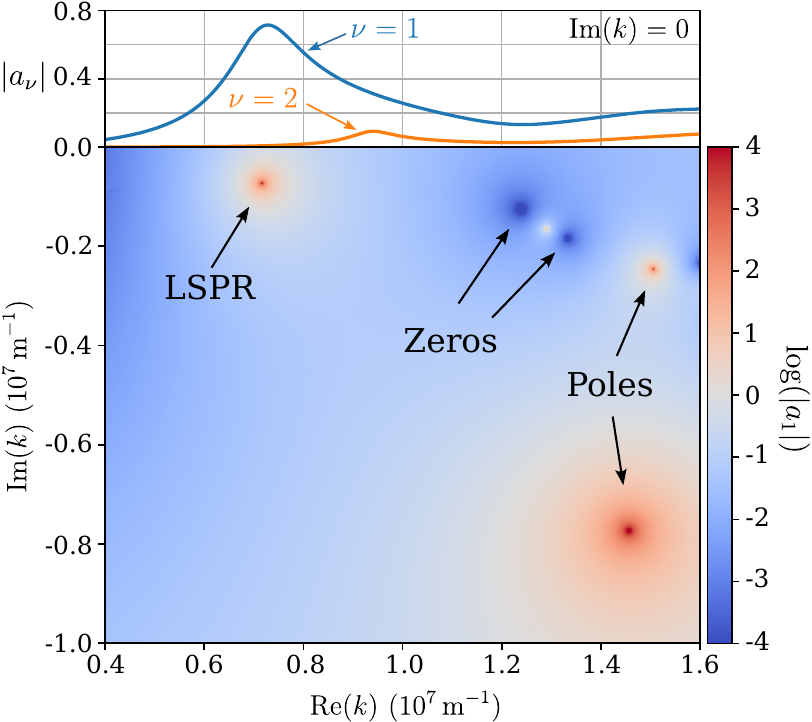}
	\caption{Heat map of $\log(|a_1|)$ for different complex wavenumbers $k = \real(k) + i\imag(k)$. Dark red regions correspond to poles and dark blue regions correspond to zeros. Cross sections of $|a_1|$ and $|a_2|$ for $\imag(k_p) =0$ are also shown. }
	\label{fig:poleszeros}
\end{figure}
The sensor design problem is equivalent to finding nanostructures that optimize $\eta_{b}$. In this work, we limit our study to two-layer (core-shell) nanospheres situated in water and consisting of a silica core and gold shell. The refractive index of water $n_b$ was modeled using a standard dataset, which was interpolated to values at arbitrary wavelengths \cite{Hale:73}. The core refractive index $n_c$ was described by a standard Sellmeier equation \cite{Malitson:65}, while the shell refractive index $n_s$ was modelled using a Sommerfield-Drude model that incorporates inter-band transitions \cite{Rakic:98}. With these material properties, the core radius $r_c$ and shell thickness $d_s$ remain as free parameters and define the design space for the nanoparticle. A schematic diagram of the nanoparticle is shown in Fig.~\ref{fig:cs}. Generalized Mie coefficients $a_\nu$ for a multi-layered sphere can be calculated using a recursive extension to Mie theory and can be used to calculate other scattering functions of interest \cite{kristensson2016scattering}. Fig.~\ref{fig:cs} shows the extinction, scattering and absorption cross sections (denoted collectively by $\sigma$) as a function of $k$ for a coated sphere with $r_c = 60\,\mathrm{nm}$ and $d_s = 10\,\mathrm{nm}$. A clear peak in all three cross sections is observable at $k \approx 0.7\times 10^7\,\mathrm{m}^{-1}$, which corresponds to a vacuum wavelength of $\approx 880\,\mathrm{nm}$ ($\approx 660\,\mathrm{nm}$ in water). This peak is characteristic of a localized surface plasmon resonance (LSPR) in gold \cite{Amendola_2017} and originates mathematically from the particle's electric dipole term $a_1$, as can be seen in the top panel of Fig.~\ref{fig:poleszeros}, which shows $|a_1|$ and $|a_2|$ over the same range of wavenumbers. A secondary, smaller peak in the extinction cross section can be observed at $k \approx 0.9\times 10^7\,\mathrm{m}^{-1}$, which is associated with a resonance in the electric quadrapole term $a_2$. The main panel of Fig.~\ref{fig:poleszeros} shows the variation of $\log(|a_1|)$ over the complex $k$ plane for the same nanosphere. The primary LSPR in Fig.~\ref{fig:cs} appears in Fig.~\ref{fig:poleszeros} as a red peak indicated near the top left corner of the main plot. The location of this pole can be found numerically by solving $1/a_1 = 0$, which we found gave $k_p \approx (0.71 - 0.07i)\times 10^7\,\mathrm{m}^{-1}$, corresponding to a resonance with $Q \approx 4.8$. The other red peaks {(`Poles')} on the right side of Fig.~\ref{fig:poleszeros} correspond to broader resonances associated with larger values of $\real(k)$ outside of the visible region. Additionally, a pair of zeros can be observed near the top right corner of the plot, which contribute to a local minimum in the real spectrum of $|a_1|$ and the extinction cross section near $k \approx 1.2\times 10^7\,\mathrm{m}^{-1}$. 

 \begin{figure}[t]
	\centering
\includegraphics[width=\columnwidth]{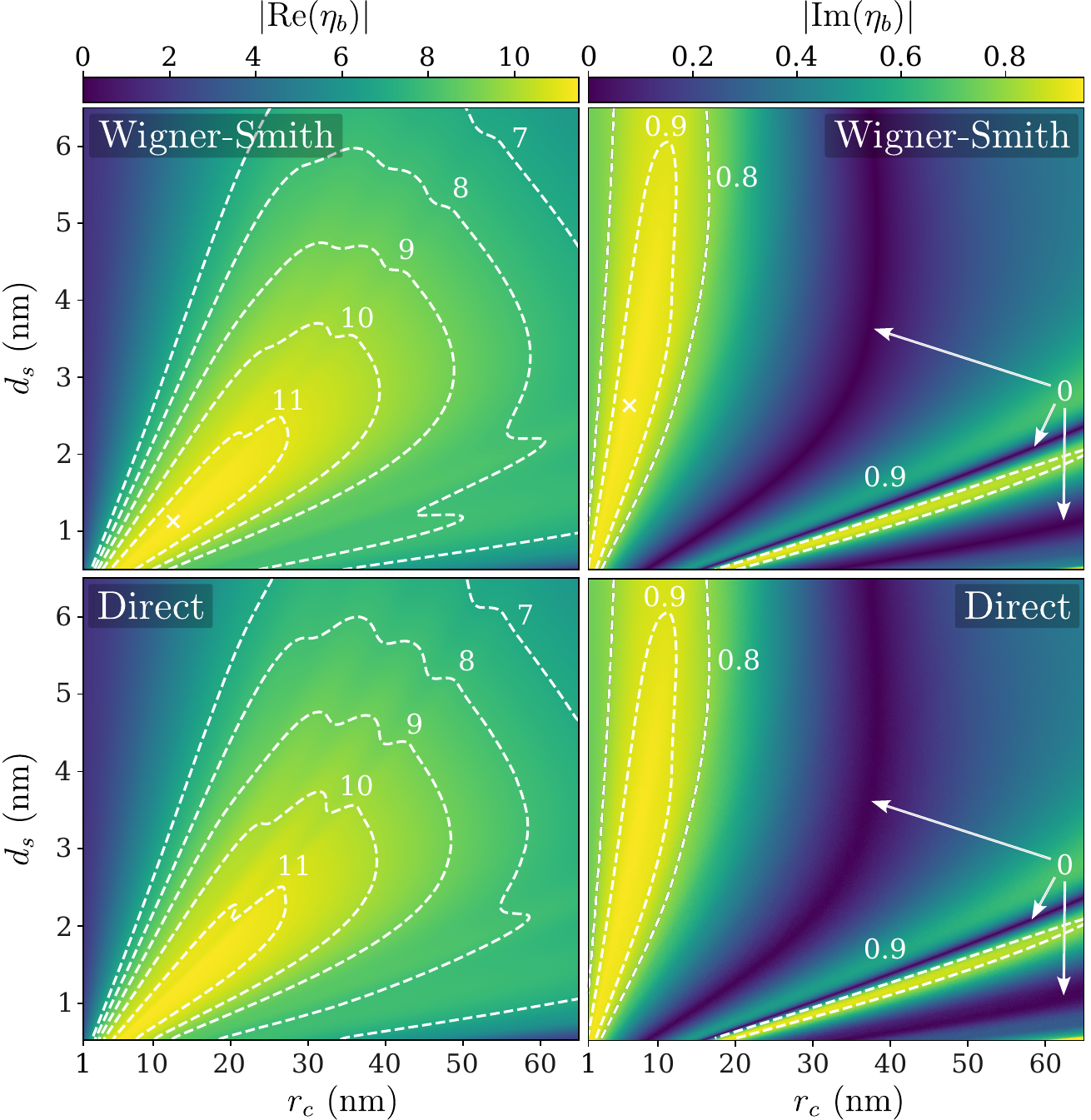}
	\caption{Heat maps of $|\real(\eta_{b})|$ (left column) and $|\imag(\eta_{b})|$ (right column) for the  LSPR indicated in Fig. \ref{fig:poleszeros} as a function of core radius $r_c$ and shell thickness $d_s$, calculated using both the Wigner-Smith based approach (top row, `Wigner-Smith') and by perturbing the nanoparticle manually as described in the main text (bottom row, `Direct'). 
}
	\label{fig:fom}
\end{figure}

Using the identified LSPR, we now examine how $\eta_{b}$ varies for different values of $r_c$ and $d_s$. Fig.~\ref{fig:fom} shows plots of $|\real(\eta_{b})|$ and $|\imag(\eta_b)|$ for $1\,\mathrm{nm}\leq r_c \leq 65\,\mathrm{nm}$ and $0.5\,\mathrm{nm} \leq d_s \leq 6.5\,\mathrm{nm}$, calculated by finding the residue in the right hand side of Eq.~(\ref{eq:fom}) (top row, or `Wigner-Smith'), as well as by numerically computing $\Delta k_p$ by manually perturbing $n_b$ (bottom row, or `Direct'). {The `Direct' method here serves as a ground truth to validate the Wigner-Smith approach.} Overall, comparing the two rows of Fig.~\ref{fig:fom}, we see that the Wigner-Smith theory agrees well with the direct numerical calculations. Minor discrepancies between the results of the two methods can be seen, particularly for the $|\real(\eta_b)|$ data, such as around the $|\real(\eta_{b})| = 8$ contour in the bottom right region of the plots, which is visibly smoother for the data calculated with the Wigner-Smith approach. These errors may have arisen from, for example, numerical inaccuracies resulting from finite difference approximations made in computing the integrals and derivatives associated with the two methods. We note, however, that the value of $\eta_{b}$ changes slowly in this region, and the numerical differences between the two methods are thus small. 

With regard to the design of the plasmonic nanoparticle, we note first that for $|\real(\eta_b)|$ there is a bright region in the lower left corner of the plot, with a peak value of $|\real(\eta_{b})|\approx 11.43$ occurring at $r_c \approx 13.3\,\mathrm{nm}$ and ${d_s} \approx 1.2\, \mathrm{nm}$ (marked with a cross in the top left panel). We found that the resonance width generally decreased with decreasing particle size, which tends to increase $\eta_b$. Larger particles, however, were found to be more sensitive to $n_b$ perturbations, resulting in a trade-off between two opposing factors that yielded a maximum value of $|\real(\eta_{b})|$ at a non-zero particle size. At this point it should be clearly noted that, particularly at the lower limits of the sizes considered, quantum and size-dependent effects can play an important role in governing the nanoparticle's resonant response \cite{Kreibig_1974}. These effects, however, are neglected in this work for simplicity. Observing the data for $|\imag(\eta_b)|$, we see that the sensitivity of the resonance width is generally much smaller than that of the central frequency, attaining a maximum value of only $|\imag(\eta_b)|\approx 0.94$ at $r_c \approx 6.3\,\mathrm{nm}$ and $d_s\approx 2.7\,\mathrm{nm}$ (marked with a cross in the top right panel). This indicates that monitoring shifts in the resonant frequency is preferable as a sensing mechanism to observing mode broadening. For some nanoparticle designs, the resonance width is in fact completely stable to perturbations. This is the case for parameters that lie on the indicated dark bands in Fig.~\ref{fig:fom}.

 \begin{figure}[t]
	\centering
\includegraphics[width=\columnwidth]{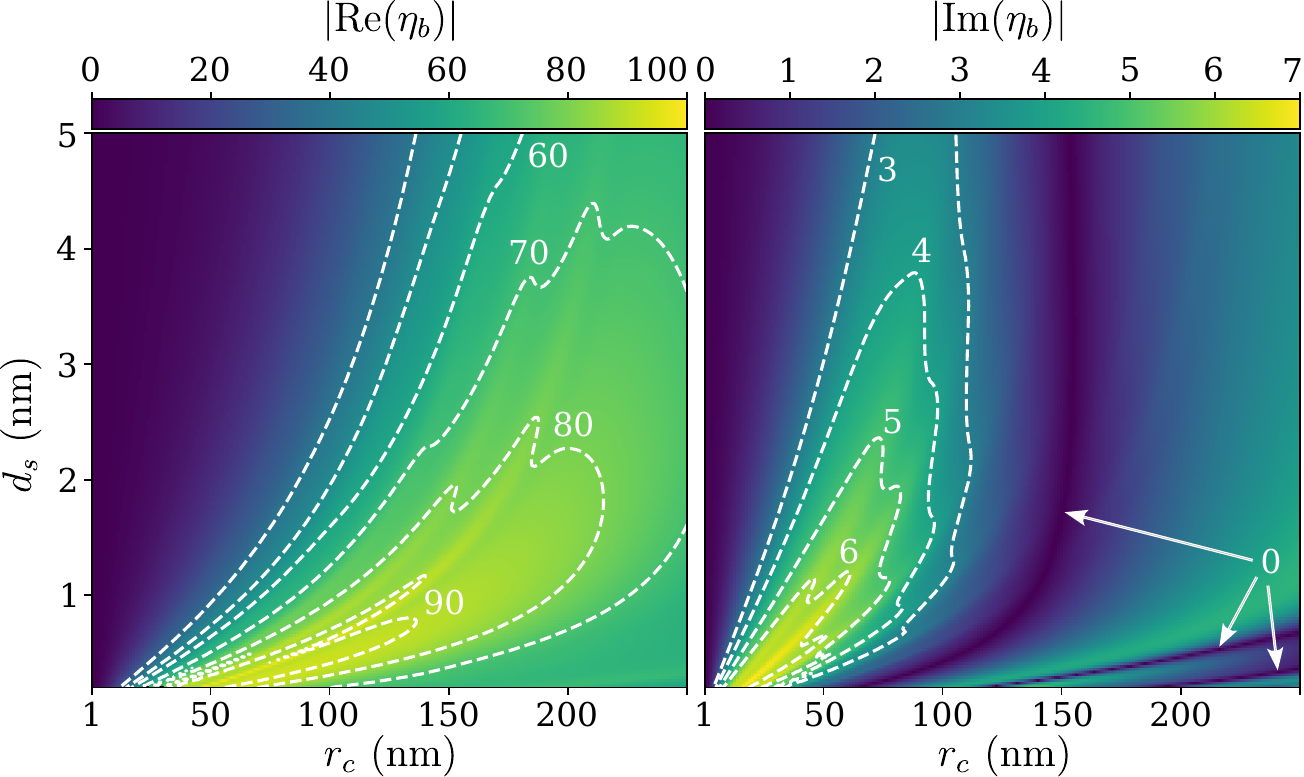}
	\caption{Heat maps of $|\real(\eta_{b})|$ (left panel) and $|\imag(\eta_{b})|$ (right panel) for the lower frequency zero shown in Fig. \ref{fig:poleszeros} as a function of core radius $r_c$ and shell thickness $d_s$.}
	\label{fig:fom-zero}
\end{figure}
As a final application of our theory, we now study the sensitivity of the nanoparticle's complex zeros to background refractive index perturbations. Complex zeros and their associated modes are linked to various interesting phenomena. For example, by correctly tailoring the temporal profile of the incident field in accordance with the mode profile associated with the zero, it is possible to achieve virtual perfect absorption, whereby the scatterer behaves as a perfect absorber \cite{Krasnok2019}. In other cases, it is possible to exploit the interference of multiple incident beams at the frequency of the zero to achieve perfect coherent absorption \cite{Baranov2017}. Although scattering zeros do not necessarily manifest as starkly in the spectra of traditional scattering quantities as poles, their positions in the complex plane can be very sensitive to perturbations, suggesting potential application in enhanced sensing technologies \cite{del2021demand}.

As discussed in section \ref{sec:alternate}, shifts in zeros can be tracked by tracking the poles of the multiplicative inverse of the operator in which they manifest.
Returning to Fig.~\ref{fig:poleszeros}, we recall the observed pair of zeros and now focus our attention to the one with smaller wavenumber. For this zero, we repeated the same analysis as was performed for the LSPR to assess its sensitivity to perturbations in $n_b$. We again used Eq.~(\ref{eq:fom}) to calculate $\eta_b$, but now took $L_{n_b}$ to be the logarithmic derivative operator associated with $M = 1/a_n$. Specifically, $\eta_{b}$ was calculated for core and shell sizes in the ranges $1\,\mathrm{nm} \leq r_c \leq 250\,\mathrm{nm}$ and $0.2\,\mathrm{nm} \leq d_s \leq 5\,\mathrm{nm}$. Fig.~\ref{fig:fom-zero} shows the calculated values of $|\real(\eta_{b})|$ and $|\imag(\eta_b)|$, presented in the same style as in Fig.~\ref{fig:fom}. Observing first the left panel of Fig.~\ref{fig:fom-zero}, we see that, although qualitatively similar to Fig. \ref{fig:fom}, $|\real(\eta_{b})|$ attains much larger values for the zero than for the LSPR, peaking at $|\real(\eta_{b})|\approx 100$ for $r_c \approx 40\,\mathrm{nm}$ and ${d_s} \approx 0.2\,\mathrm{nm}$: an order of magnitude greater than the maximum value for the LSPR. This difference results from a much larger sensitivity to perturbations for the zero than for the LSPR, as it can be seen from Fig.~\ref{fig:poleszeros} that the LSPR and zero both have similar imaginary parts. A similar conclusion can be reached for $|\imag(\eta_b)|$, which for the zero peaks at $|\imag(\eta_b)|\approx 7$ for $r_c \approx 26\,\mathrm{nm}$ and $d_s \approx0.2\,\mathrm{nm}$. Some care must be taken in making direct comparisons between Figs.~\ref{fig:fom} and \ref{fig:fom-zero}, as the LSPR and zero occur at different complex wavenumbers. Though our work serves as a proof of principle, using our approach one could perform a more thorough comparative analysis of the sensitivities of different poles and zeros throughout the complex plane, employing alternative optimization metrics as required. It is also conceptually straightforward to extend the analysis to different types of perturbations, such as the thickness of a particular layer, or, given a suitable model, anisotropic changes to the system parameters.  

\section{Conclusion}\label{sec:conclusion}
In summary, we have given a detailed account of how the generalized Wigner-Smith theory of resonance shifts as described by Eq.~(\ref{eq:ws-formula}) is related to the generalized cavity perturbation theory in which one takes unconjugated products of the system's internal fields, therefore extending the results of Ref.~\cite{Byrnes2024b} to resonances with arbitrary $Q$ factors. Furthermore, we have presented additional insight into how the Wigner-Smith theory generalizes to open scattering geometries and how it can be applied to poles and zeros of more general scattering functions, such as scattering coefficients and cross sections. We applied our theory to several practical examples, including scattering by a homogeneous sphere and the inverse design of a silica-gold core-shell optical nanosensor. Finally, we demonstrated the enhanced sensitivity of scattering zeros to system perturbations. We envisage that our work will serve as a useful tool in the analysis and design of future resonance-based systems and technologies.

\begin{acknowledgments}
N.B. was supported by Singapore Ministry of Education Academic Research Fund (Tier 1) Grant RG66/23. M.R.F. was supported by funding from the Institute for Digital
Molecular Analytics and Science (IDMxS) under the Singapore Ministry of Education Research Centres of Excellence scheme and by Nanyang Technological University Grant SUG:022824-00001.
\end{acknowledgments}

\providecommand{\noopsort}[1]{}\providecommand{\singleletter}[1]{#1}%

\end{document}